%% file: conf_0211.tex
\documentclass[11pt]{article}
\usepackage{graphicx,amsmath}

\input{pubboard/babarsym}
\input{symbols}

\def\pipi     {\ensuremath{\pi\pi}}

\def\fish    {\ensuremath{\cal F}}
\def\cerenkov{$\check{\rm C}{\rm erenkov}$}

\newcommand{\BABARPubYear}    {02}

\newcommand{\BABARConfNumber} {11}
\newcommand{\SLACPubNumber} {9270}

\long\def\inst#1{\par\nobreak\kern 4pt\nobreak
    {\it #1}\par\vskip 10pt plus 3pt minus 3pt}

\begin{document}
\pagestyle{empty}

\begin{flushright}
\babar-CONF-\BABARPubYear/\BABARConfNumber \\
SLAC-PUB-\SLACPubNumber \\
Jun 2002 \\
\end{flushright}

\par\vskip 3cm

\begin{center}
{
\Large \bf
 Measurements of Charmless Two-Body Charged \boldmath{$\B$} Decays with  Neutral Pions and Kaons
}
\end{center}
\bigskip

\begin{center}
\large The \babar\ Collaboration\\
\mbox{ }\\
\today
\end{center}
\bigskip \bigskip

\begin{center}
\large \bf Abstract
\end{center}
We present preliminary results of the analyses of  $\B \to h \piz$ and 
$\B \to h  \Kz $ decays  (with $ h$ = $\pi^{\pm}, K^{\pm}$) from  a sample 
of approximately $60$ million \BB\ pairs collected by the \babar\ detector 
at the \pep2\ asymmetric-energy $B$ Factory at SLAC.  
We find evidence for a signal in $\Bp \to \pip \piz$, and we measure the 
branching fraction 
\[
\BR (\B^+ \to \pi^+ \pi^0) =  (4.1 _{-1.0}^{+1.1}\pm 0.8) \times 10^{-6}.
\]
We also measure the following branching ratios and charge asymmetries:
$\BR(\B^+ \to K^+ \pi^0) = (11.1 _{-1.2}^{+1.3} \pm 1.0) \times
10^{-6}$,  $\BR(\B^+ \to \pi^+ \Kz) = (17.5^{+1.8} _{-1.7} \pm 1.3)
\times  10^{-6}$,  $\BR(\B^+ \to K^+ \Kzb) < 1.3 \times 10^{-6}$
($90\%$ CL), $\acp_{\pi^+ \pi^0} = -0.02^{+0.27}_{-0.26} \pm 0.10$,
$\acp_{K^+ \pi^0} = 0.00\pm 0.11 \pm 0.02$,  $\acp_{\pi^+ \Kz} =
-0.17\pm 0.10 \pm 0.02$, where the errors are  statistical and
systematic, respectively.

\vfill
\begin{center}
 Presented at the Conference on Flavor Physics and \CP\ Violation (FPCP)
\\ 5/16--5/18/2002, Philadelphia, PA, USA
\end{center}

\vspace{1.0cm}
\begin{center}
{\em Stanford Linear Accelerator Center, Stanford University,
Stanford, CA 94309} \\ \vspace{0.1cm}\hrule\vspace{0.1cm} Work
supported in part by Department of Energy contract DE-AC03-76SF00515.
\end{center}

\newpage
\pagestyle{plain}

\input  pubboard/authors_conf02011.tex \setcounter{footnote}{0}
\section {Introduction}

The study of $B$ meson decays into charmless hadronic final states
plays an important role in the understanding of \CP\ violation in the
$B$ system.  Measurements of the \CP-violating asymmetry in the $\pip\pim\ $
decay mode can provide  information on the angle $\alpha$  of the
Unitarity Triangle.  However, in contrast to the theoretically clean
determination of the angle $\beta$ in $B$ decays to charmonium final
states \cite{sin2betaBaBar,sin2betaBelle} the  extraction
of $\alpha$ in $\pip\pim$ decay is complicated by the interference of
$b\to uW^-$ tree and $b\to dg$ penguin amplitudes. Since these
amplitudes have similar magnitude but carry different weak phases, 
additional measurements of the isospin-related decays\footnote{Charge
conjugate modes are assumed throughout this paper.}, $\Bp \to \pip  \piz$
and $\Bz \to \piz \piz$, are required to provide a means of measuring
$\alpha$~\cite{ref:gronau2}.  The measurement of the
branching ratio of the $\Bp \to \pip \piz$ decay is, in fact, a
crucial ingredient, since it is a pure tree amplitude decay to a very
good approximation. Therefore, in this channel direct \CP\ violation, detected as a
charge asymmetry (\acp), is  expected to be zero. 
 Moreover, measurements  of $\B \to K\pi$  decays are  interesting since
phenomenological models have been proposed for extracting the weak phase $\gamma$ with
a global fit to the observables \cite{beneke,ciuchini,Keum:2000wi}.  We also
 present here  an analysis of the $\Bp \to \pip\Kz$ and $\B^+
\to  K^+ \Kzb$ decays.  The \babar\ collaboration has previously
published \cite{ourPRL} measurements of the branching fractions for $B$ mesons decaying into  $\Kp
\piz$ and  $\B^+ \to  \pi^+ \Kz$, but no significant signals were seen for  $\Bp \to \pip \piz$ and $\B^+ \to  K^+  \Kzb$ decays.  The results reported here are an update of these
published analyses.

\section {Data Sample}

The data used in these analyses were collected with the \babar\ detector
at the \pep2\ $\epem$ storage ring during the years 2000 and 2001.  
The sample corresponds to an integrated luminosity of  about $54\invfb$ accumulated  near the
\FourS\ resonance (``on-resonance'') and about $5 \invfb$ accumulated at a 
center-of-mass (CM) energy about $40\mev$ below the \FourS\ resonance
(``off-resonance''), which are used for continuum background studies.
The on-resonance sample corresponds to $(60.2 \pm 0.7)\times 10^6$ 
\BB\ pairs.  The collider is operated with asymmetric beam energies, 
producing a boost ($\beta\gamma = 0.55$) of the \FourS\ along the 
collision axis.  The boost increases the momentum range of 
two-body $B$ decay products from a narrow distribution centered near 
$2.6\gevc$ in the CM to a broad distribution extending from $1.7$ to $4.3\gevc$.

\babar\ is a solenoidal detector optimized for the  asymmetric beam
configuration at PEP-II and is described in detail in Ref.~\cite{babarnim}.
Charged  particle (track) momenta are measured in a tracking system
consisting of a  5-layer, double-sided, silicon vertex tracker and a
40-layer drift chamber  filled with a gas mixture of helium and
isobutane, both operating within a  $1.5\,{\rm T}$ superconducting
solenoidal magnet.  Photon candidates  are selected as local maxima of
deposited energy   in an electromagnetic calorimeter (EMC) consisting
of 6580 CsI(Tl) crystals arranged in  barrel and forward endcap
subdetectors.   In this analysis, tracks are identified as pions or
kaons by the \cerenkov\ angle  $\theta_c$ measured by a detector of
internally reflected Cherenkov light (DIRC).  The DIRC system is a
unique type of Cherenkov detector that relies on total internal
reflection within the radiating volumes (quartz bars) to deliver the Cherenkov light outside
the tracking and magnetic volumes, where the Cherenkov ring is imaged
by an array of $\sim 11000$ photomultiplier tubes.
  
\section {Event Selection,  \boldmath{$\piz$}  and  \boldmath{$\Kz$} 
Reconstruction}

Hadronic events are selected based on track multiplicity and event
topology.  Backgrounds from non-hadronic events are reduced by
requiring the ratio of Fox-Wolfram moments, $H_2/H_0$ \cite{fox}, to be
less than $0.95$ and the sphericity \cite{spheric} of the event to be
greater than $0.01$.

Candidate \piz\ mesons  are reconstructed as pairs of photons
with an invariant mass within $3 \sigma$ of the nominal \piz\
mass~\cite{PDG}, where the resolution $\sigma$ is about 8 \mevcc.  Photon candidates are selected as showers in the EMC that  have
the expected lateral shape, are not matched to a track, and have a
minimum energy of 30 \mev.  The \piz\ candidates are then kinematically
fitted with their mass constrained to the \piz\ nominal mass.

\Kz\ mesons are detected in the mode $\Kz\to \KS\to \pip\pim$ and
are reconstructed
from pairs of oppositely charged tracks that form a
well-measured vertex and have an invariant mass within $11.2\mevcc$
(which corresponds to $3.5\sigma$) of the  nominal \KS\ mass~\cite{PDG}.  The measured
proper decay time of the \KS\ candidate is required to exceed five
times its uncertainty.

\section {\boldmath{$B$} Reconstruction }
\label{brecosec}

$B$ meson candidates are reconstructed by combining a \piz  or a \KS
candidate with a track $h$. The kinematic constraints provided by the
\FourS\ initial state and  knowledge of the beam
energies are exploited to efficiently identify $B$ candidates.   We
define  a beam-energy substituted mass  $\mes = \sqrt{E^2_{\rm
b}-\mathbf{p}_B^2}$,  where $E_{\rm b} =(s/2 + \mathbf{p}_i
\cdot\mathbf{p}_B)/E_i$, $\sqrt{s}$ and $E_i$ are the total energies
of the \epem\ system in the CM and lab frames, respectively, and
$\mathbf{p}_i$ and $\mathbf{p}_B$ are  the momentum vectors in the lab
frame of the \epem\ system and the $B$ candidate, respectively. 
 An  additional kinematic parameter $\Delta E$  is defined as the
difference between the energy of the $B$ candidate and half the energy
of the \epem\ system, computed in the CM system. 
 The \mes\ resolution is  dominated by the beam energy spread, while  for  $\Delta E$  the main contribution comes from the measurement of particle energies in the detector. These two variables are therefore substantially uncorrelated.

 However, in the $h \piz$ (with $ h$ = $\pi^{\pm}, K^{\pm}$) final states  both the $\Delta E$ and $\mes$  distributions have  a tail due to  imperfect containment of the electromagnetic showers initiated by the $\piz$. In this case only, in order to reduce  this source of correlation and to slightly improve the resolution,  we fit  the $B$ candidate with the energy constrained to the CM beam energy in the two cases of  kaon and pion  mass hypothesis for the track $h$. For the $h \piz$ decay  the energy-constrained mass resolution  is then found  to be about 3 \mevcc  from the core Gaussian width of a Crystal Ball\footnote{ A core Gaussian with a power law to describe a tail at negative values is  called the Crystal Ball function \cite{cryball}. }  fit to Monte Carlo simulated signal   events. For the $h \KS$ decay the \mes resolution is found to be $2.5 \mevcc$ from a Gaussian fit.  For both decay topologies the signal Monte Carlo resolutions are  validated by comparing data and Monte Carlo resolutions for decays into open charm final states with large branching fractions, such as  $\Bm \to D^0 \rho^{-}$, (with $\rho^{-} \to \pim \piz$ and $D^0 \rightarrow K^- \pi^+$) for the $h \piz$ analysis, and $\Bm \to D^0 \pi^{-}$ ($D^0 \rightarrow K^- \pi^+$) for the $h \KS$ analysis.

  The $\Delta E$ variable is  evaluated assuming the pion mass hypothesis for the track $h$. Its distribution for the signal $\pip \piz$ events is described by  a Crystal Ball function centered near zero. Since the $\Delta E$ distribution has a mean that depends  on the track $h$ momentum in the lab frame in the case of signal $\Kp \piz$ events, we also calculate $\Delta E$ with the kaon mass hypothesis ($\Delta E (K)$) for those events. We empirically find that   its distribution is  described better by  a sum of two Gaussians with different mean values. For $h \KS$ signal events the $\Delta E$ distribution is parametrized as  a  sum of two Gaussians centered near zero with the core Gaussian accounting for 95\% of the  events, taking into account the momentum dependance  for signal $\Bp \to \Kp \KS$.   Based on Monte Carlo simulated $ \Bp \to \pip \piz$ and $ \Bp \to \pip \KS$ events, we estimate the resolution on $\Delta E$ for  the core Gaussian width to be about $40$ \mev and $26$ \mev, respectively. Candidates are selected in the range  $5.2<\mes<5.3\gevcc$. Different  requirements  on $\Delta E$  specific to each analysis are then applied.

\section { Background Rejection}
 The dominant background to these channels is from random combinations of a true \piz (\KS)  with a track, produced in  $\epem\to \qqbar$  continuum events (where $q=u$, $d$,$s$, or $c$). Another source of background originates from  $B$ decays into three (or more) light mesons. 
Detailed Monte Carlo simulation, off-resonance, and on-resonance data  are used to study backgrounds.  For this study we select on-resonance data in $\Delta E$   sideband regions  defined by the ranges $0.20 < |\Delta E| < 0.45\gev$ for  $h \piz$,  and  $-0.305 < \Delta E < -0.115 \gev$ plus $0.075 < \Delta E < 0.265 \gev$ for $h \KS$.  

 In the CM frame the continuum   background typically exhibits a
two-jet structure, in contrast to the isotropic decay of $\BB$ pairs
produced in \FourS decays.  We exploit the topology difference between
signal and background by making use of two event-shape quantities.

The first variable is the angle  $\thsph$ between the sphericity axes
of the $B$ candidate and of the  remaining tracks and photons in the
event.  The distribution of $|\cos\thsph|$ in the CM frame is strongly
peaked near $1$ for continuum events and is approximately uniform for
\BB\ events.  We require $|\cos\thsph| < 0.8$ in the $h \piz$ analysis,
but, given the lower level of background, only $|\cos\thsph| < 0.9$
in the $h \KS$ analysis.

The second quantity is a Fisher discriminant ${\cal F}$ \cite{ourPRL} 
 constructed from the scalar sum of the CM momenta of all tracks and photons
(excluding the $B$ candidate decay products) flowing into nine
concentric cones centered on the thrust axis of the $B$ candidate.
Each cone subtends an angle of $10^\circ$ and is folded to combine the
forward and backward intervals.  Monte Carlo samples are used to
obtain the values of the Fisher coefficients, which are determined by
maximizing the statistical separation between signal and background
events. No requirement is applied on ${\cal F}$; instead the distributions for
signal and background events  are included in a maximum likelihood fit as
described in the next section.

  On the other hand,  $B$ background events tend to peak in \mes, as do signal
events, but have more  negative $\Delta E$ values. They are particularly harmful for the $h \piz$ analysis given the poorer $\Delta E$ resolution.  We use data in the negative $\Delta E$ sideband region to
estimate the magnitude of this background and Monte Carlo
techniques to choose a $\Delta E$ requirement that reduces
this background to a negligible level. We finally  require $-0.11 < \Delta E  < 0.15 \gev$  for $h \piz$ and  $-0.115 < \Delta E < 0.075\gev$ for $h \KS$.

A total of 13661   candidates in the on-resonance data satisfy
our $ h \piz$ selection criteria and with the $h\KS$
analysis requirements  we select  10668
candidates.  These two samples enter into two separate maximum
likelihood fits.

The final  selection efficiency  $\epsilon$  is $(25.6 \pm 1.7)\%$
[$(22.5 \pm 1.5)\%$]  for  $\Bp \to \pip \piz$ [$\Bp \to \Kp
\piz$] events, while it is $(47.5 \pm  2.2 )\%$  [$(47.1 \pm 2.2)\%$]
for  $\Bp \to \pip \KS$ [$\Bp \to \Kp \KS$] events.  The errors on the efficiencies are  statistical and systematic, combined in quadrature. The dominant component is due to the imperfect knowledge of
\piz\ and \KS\ reconstruction efficiencies ($5\%$ and $3\%$ relative
errors, respectively).

\section{Signal Extraction}

For each topology ($h\piz$ and $h\KS$), an unbinned maximum likelihood fit
determines the signal and background yields $n_i$
($i=1$ to $M$, where $M$ is the total number of signal and background 
species) and charge asymmetries $\acp_i =( n_i^{-} - n_i^{+} )/(
n_i^{-} + n_i^{+})$, where $n_i^{-} \,(n_i^{+} )$ is the fitted number
of $i^{th}$ type $h^{-} \piz$ ($h^{+} \piz$) [$h^{-} \KS$ ($h^{+}
\KS$)] events.  The input variables to the fit are
\mes, $\Delta E$, \fish\ and the Cherenkov angle $\theta_c$ of the track 
from the candidate $B$ decay.  The extended likelihood function $\cal L$ 
is defined as
\begin{equation}
{\cal L}= \exp\left(-\sum_{i=1}^M n_i\right)\, \prod_{j=1}^N
\left[\sum_{i=1}^M \frac{1}{2}(1- q_j \acp_i)   n_i {\cal
P}_i\left(\vec{x}_j; \vec{\alpha}_i\right) \right]\, ,
\end{equation}
where $q_j$ is the charge of the track $h$  in the $j^{th}$ event.
The  $M$ probabilities ${\cal P}_i(\vec{x}_j;\vec{\alpha}_i)$ are
evaluated as the product of probability density functions (PDFs) for
each of the independent variables $\vec{x}_j$, given the set of
parameters $\vec{\alpha}_i$.  Monte Carlo simulation is used to
validate the assumption that the fit variables are uncorrelated.  The
exponential factor in the likelihood accounts for Poisson fluctuations
in the total number of observed events $N$.

The parameters for the  background \mes\ and $\Delta E$ PDFs are determined
from  events in the off-resonance  data and in the \mes\ sideband region,
respectively.  The \mes\ shape is parameterized by a threshold
function~\cite{argus} $f(\mes)\propto
\mes\sqrt{1-x^2}\exp[-\xi(1-x^2)]$, where $x=\mes/m_0$ and $m_0$ is
the average CM beam energy.  The background shape in  $\Delta E$ is
parameterized as a second-order polynomial.  The signal distributions
 have been already described in Sect. \ref{brecosec}.

Events from Monte Carlo simulated signal decays  and from on-resonance \mes\ sideband regions  are used to parameterize the Fisher discriminant 
PDF for signal and background events as a Gaussian and a sum of two 
Gaussians, respectively.  Alternative parameterizations for ${\cal F}$,
obtained from off-resonance data (for background) and $\Bm \to D^0 \pim$ fully 
reconstructed  decays (for signal), are used to estimate
systematic uncertainties.  The $\theta_c$ PDFs are derived from kaon
and pion tracks in the momentum range of interest from a sample of $D^{*+}\to\Dz\pip$ ($\Dz\to \Km\pip$) decays.  This
control sample is used to parameterize the $\theta_c$ resolution as a function of track polar angle.
 
The results of the fit are summarized in the first column of Table
\ref{tab:hpi0brresult}, where the statistical error for each mode
corresponds to a $68\%$ confidence level  interval and is given by the
change in signal yield $n_i$ that corresponds  to a $-2\ln{\cal L}$
increase of one unit.  We define  a signal statistical significance  as the square root of the change in $-2\ln{\cal L}$  when  the signal yield is fixed to zero.  For the $\pip \piz$ mode, we find  a $5.2 \sigma$ statistical significance for the signal.

\begin{figure}[!tbp]
\begin{center}
\begin{minipage}[h]{6.0cm}
  \includegraphics[width=6.0cm]{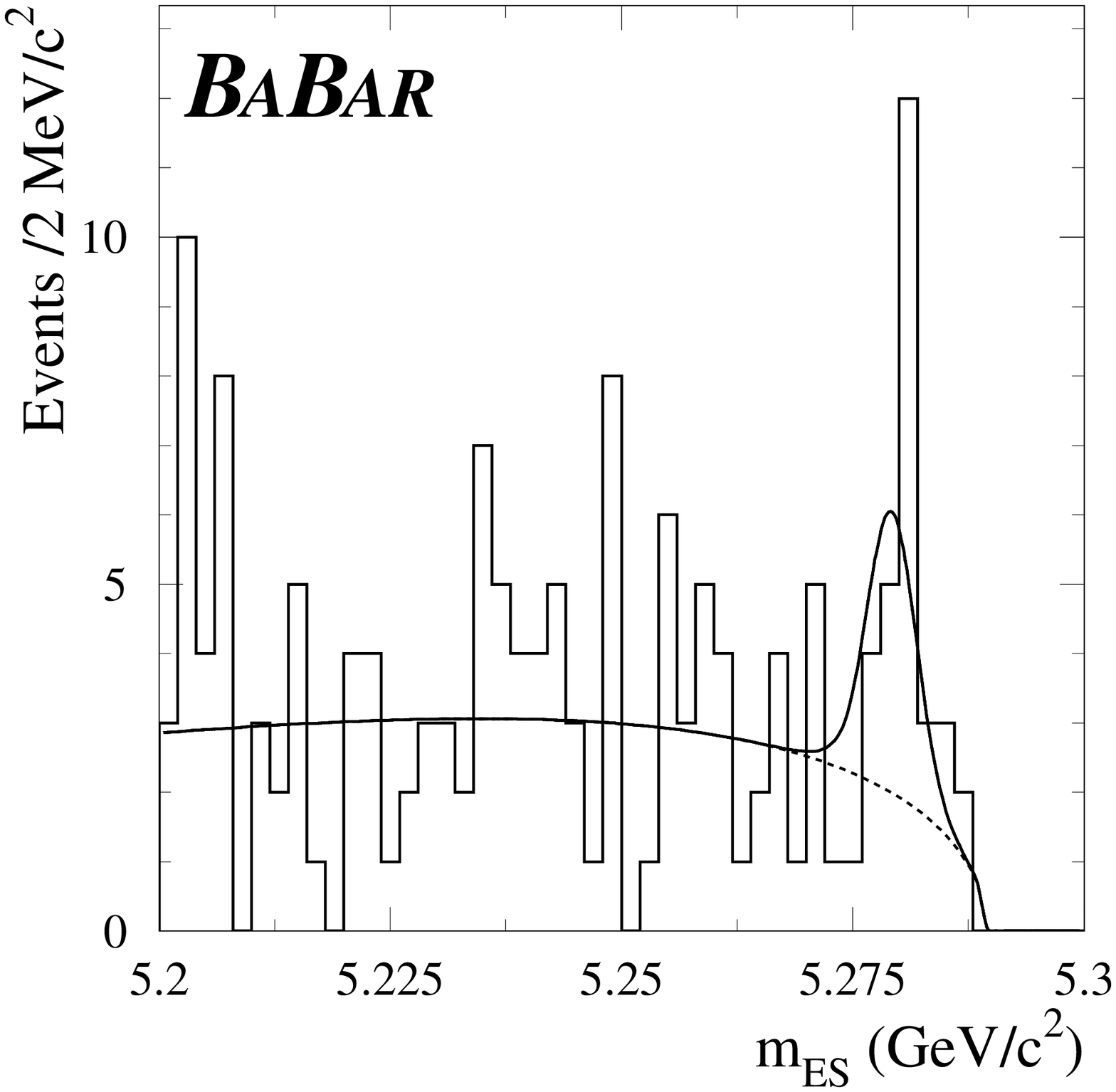}
\end{minipage}
\begin{minipage}[h]{6.0cm}
 \includegraphics[width=6.0cm]{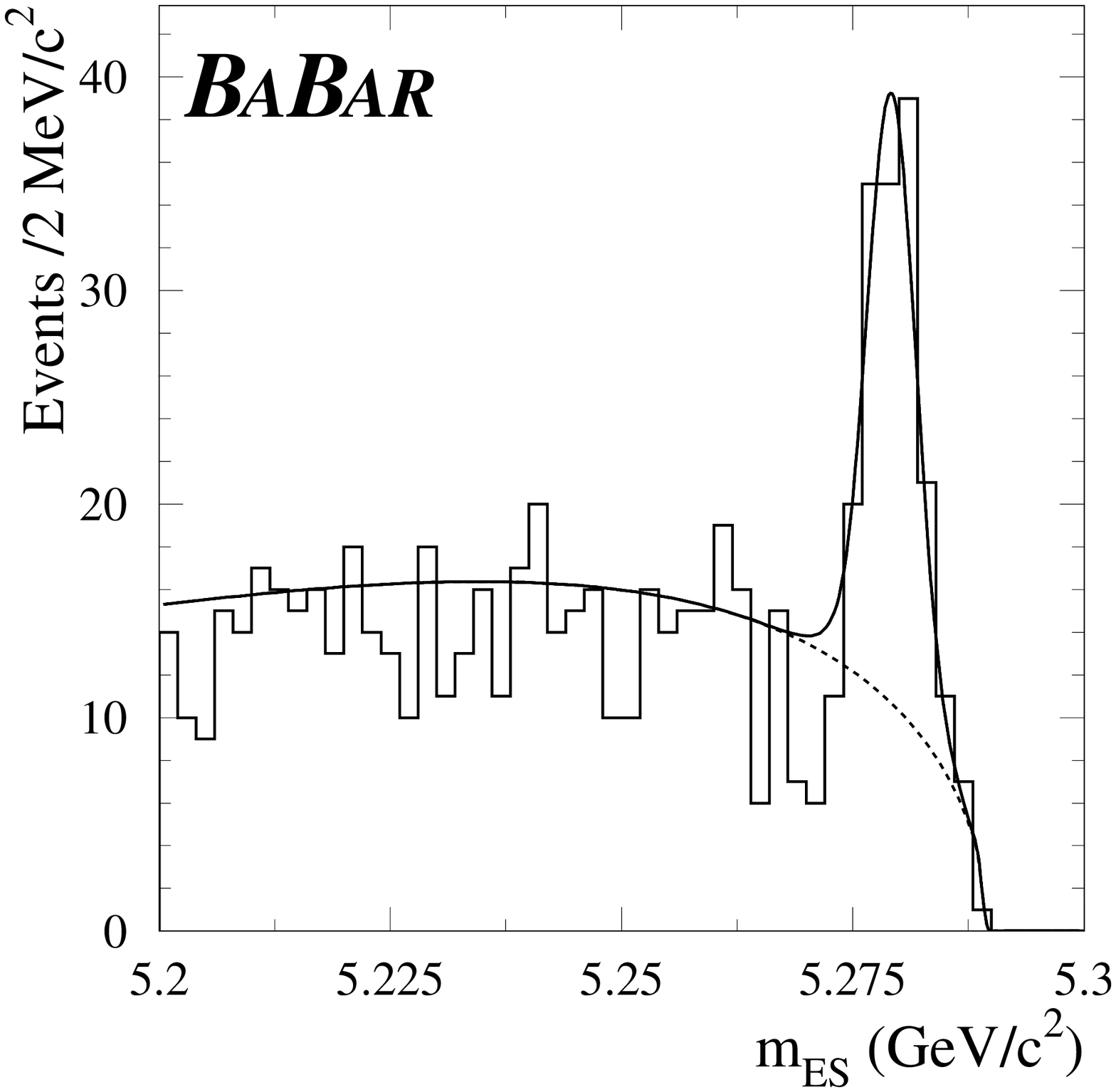}
\end{minipage}
\end{center}
\begin{center}
\begin{minipage}[h]{6.0cm}
  \includegraphics[width=6.0cm]{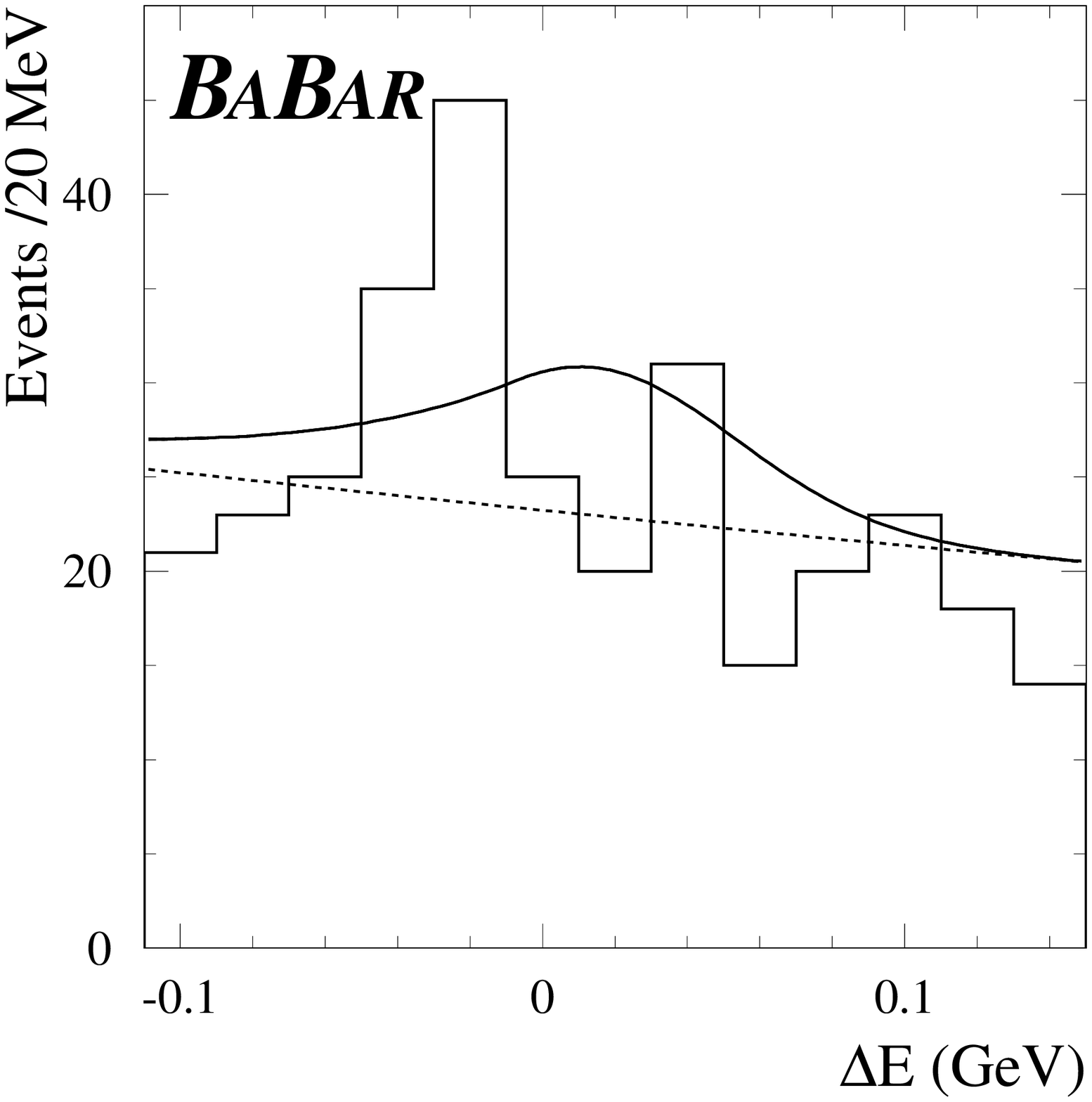}
\end{minipage}
\begin{minipage}[h]{6.0cm}
 \includegraphics[width=6.0cm]{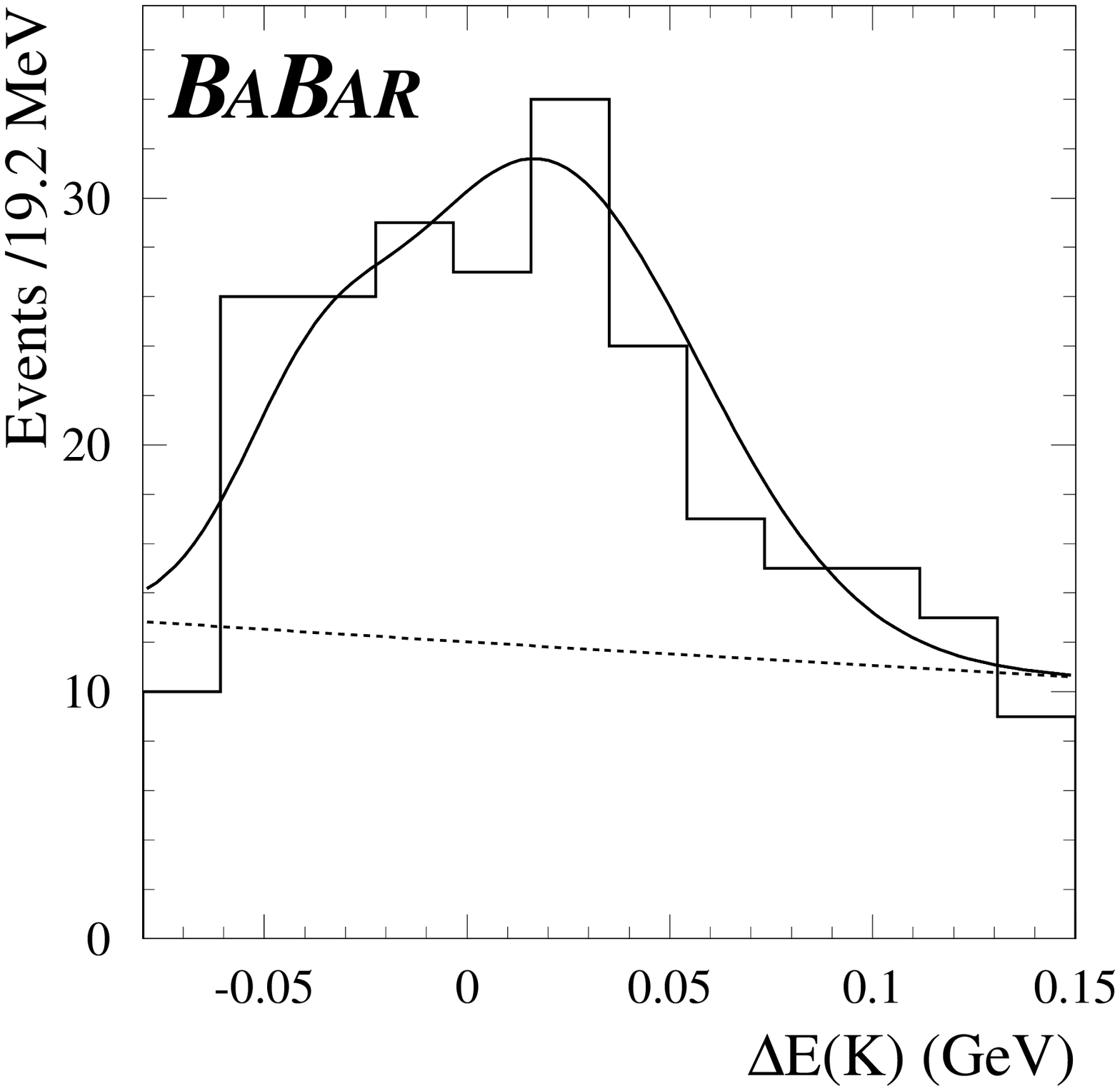}
\end{minipage}
\end{center}

\caption{Distributions of \mes  and $\Delta E$ for $\pip \piz$ events  (left) and $\Kp \piz$ events (right) after additional requirements on likelihood ratios, based on all variables except the one being plotted.  Solid curves  represent  projections of the complete maximum likelihood fit result; dotted curves represent the background contribution. }
\label{fig:prplots}
\end{figure}

\begin{figure}[!tbp]
\begin{center}
\begin{minipage}[h]{6.0cm}
  \includegraphics[width=6.0cm]{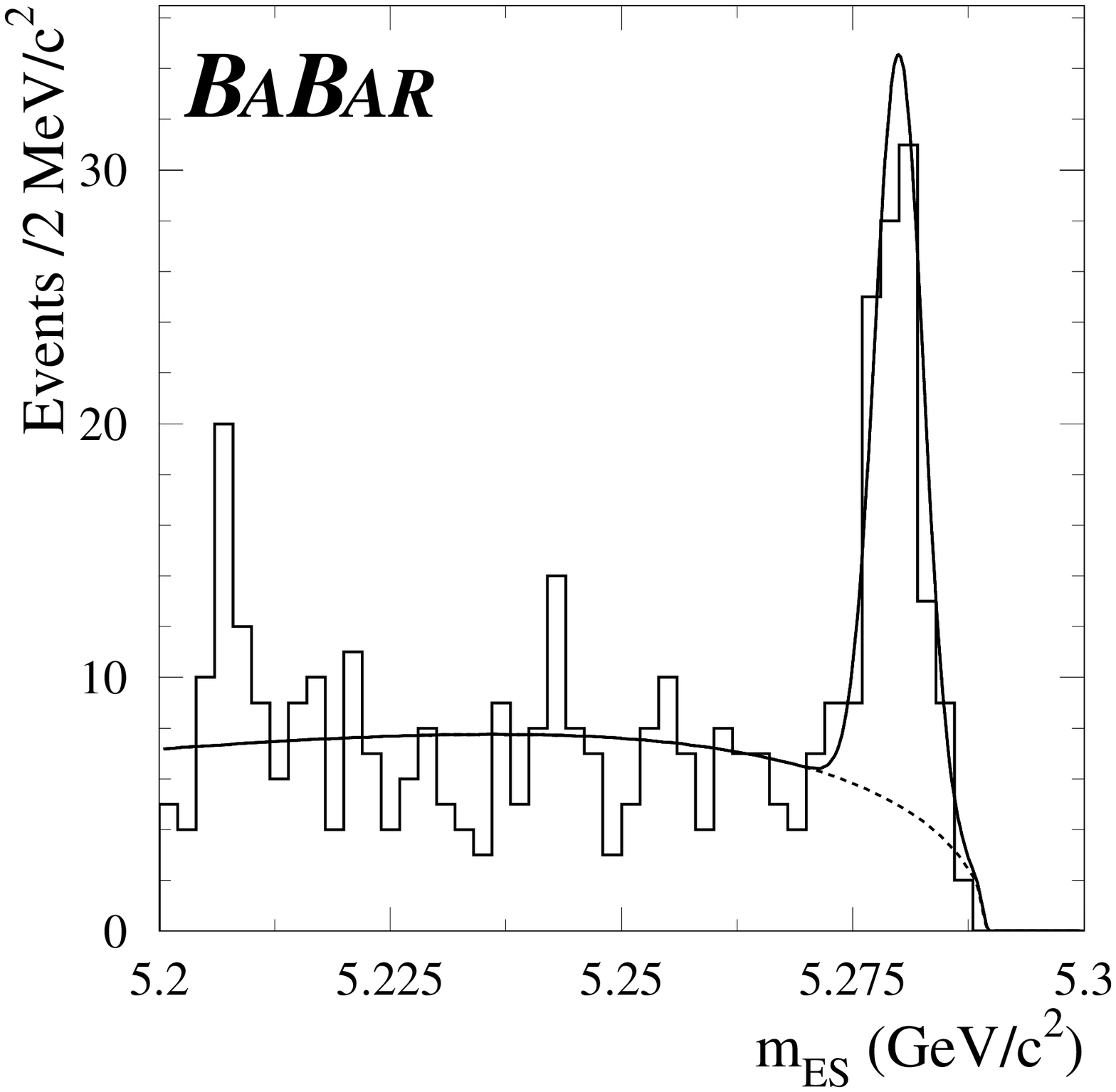}
\end{minipage}
\begin{minipage}[h]{6.0cm}
 \includegraphics[width=6.0cm]{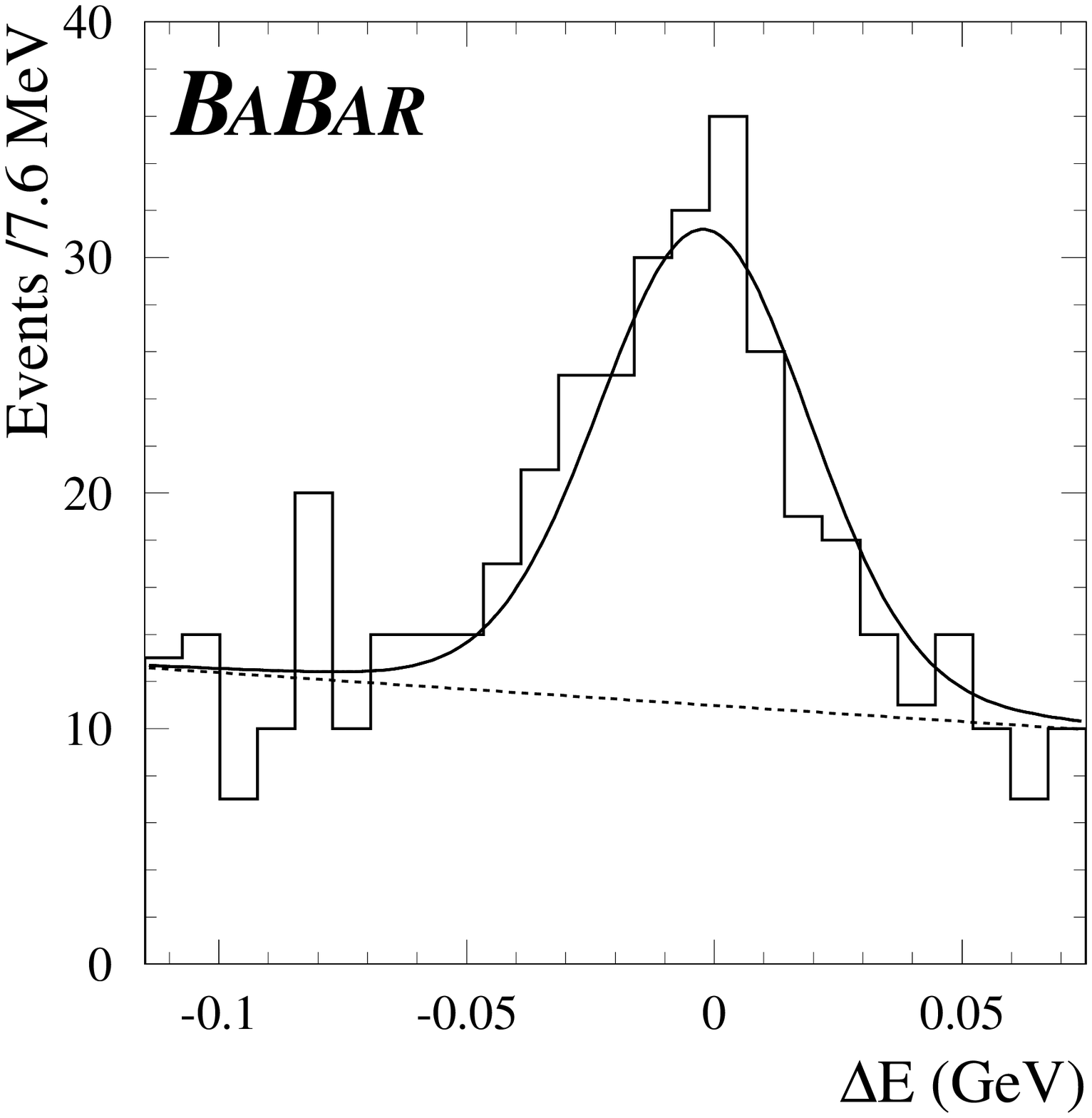}
 \end{minipage}
 \end{center}
\caption{Distributions of \mes (left) and  $\Delta E$ (right) for $\pip \KS$  events after additional requirements on likelihood ratios, based on all variables except the one being plotted. Solid curves  represent  projections of the complete maximum likelihood fit result; dotted curves represent the background contribution.   }
\label{fig:prplotshks}
\end{figure}

  In order to increase the relative  fraction of signal events of a given type for display purpose we  choose events passing requirements on likelihood ratios.    These
likelihood ratios are defined  as ${\cal R}_{\rm sig} = \sum_s  {\cal P}_s/\sum_i  {\cal P}_i$ and  ${\cal R}_k = {\cal P}_k/\sum_s {\cal P}_s$,   where $\sum_s$ denotes the sum over the probabilities for signal hypotheses only, $\sum_i$ denotes the sum over all the probabilities (signal and background),  and ${\cal P}_k$ denotes the probability for signal hypothesis $k$.  These probabilities are constructed from all the PDFs except that describing the plotted variable.    Figures \ref{fig:prplots} and  \ref{fig:prplotshks} show the distributions in \mes\  and $\Delta E $ for events passing all such selection criteria.     The likelihood fit projections, scaled by the relative efficiencies for the likelihood ratio requirements,  are overlaid on each distribution. Since the sample projections in \mes\ and $\Delta E $  are obtained with requirements on different likelihood ratios, the number of signal events appearing in the two projections are not the same.

\section{Branching Fraction Results}
The branching fractions are defined as
\begin{eqnarray}
\label{hpi0br}
{\BR}(h  \pi^0) &=&  \frac{1}{\BR (\pi^0 \to \gamma \gamma )}
\frac{n_{ h  \pi^0}}{\epsilon_{h\pi^0} \cdot N_{\BB}},
\\
\label{kspi0br}
{\BR}(h K^0) &=& \frac{1}{{\BR}(K^0 \to \KS) \cdot {\BR}(\KS \to
\pip\pim)}\frac{ n_{ h  \KS }}{\epsilon_{h\KS} \cdot N_{\BB}}\,,
\end{eqnarray}
where $n_{h \pi^0}$ ($n_{h \KS }$) is the 
 signal yield from the fit and $\epsilon_{h\piz}$ ($\epsilon_{h\KS}$) is 
the reconstruction efficiency for the mode $h\piz$ ($h\KS$) in the detected $\piz$ (\KS) decay chain.
$N_{\BB} = ( 60.2\pm 0.7)\times 10^6$ is the total number of $\BB$
pairs in our dataset.  $\BR(\pi^0 \to \gamma \gamma )$, $\BR(K^0 \to
\KS)$, and $\BR(\KS \to \pip\pim)$ are taken to be equal to
$0.98798$, $0.5$ and $0.6861$, respectively \cite{PDG}.  Implicit in
the above equations is the assumption of equal
branching fractions for $\Y4S\to\Bz\Bzb$ and $\Y4S\to\Bu\Bub$.

\begin{table}[!tbp]
\caption{Summary of   fitted signal yields, measured branching fraction \BR\, and charge asymmetries \acp .  
The first error is statistical
and the second is systematic.  For the $\Kp \Kzb$ mode we quote the $90\%$
confidence level (CL) upper limits for the signal yield and branching ratio,
and give the central values in parentheses.}
\begin{center}
\smallskip
\begin{tabular}{cccc} \hline\hline
 Mode & Signal Yield  &
 $\BR\,(10^{-6})$ &  \acp   \\ \hline
$\pip \pi^0$ & 
$62_{-16}^{+17}\pm 11$   & $4.1 _{-1.0}^{+1.1}\pm 0.8
 $ & $-0.02^{+0.27}_{-0.26} \pm 0.10$ \\
\smallskip
 $\Kp \pi^0 $    & $149\pm 17\pm 8   $  &   $11.1
_{-1.2}^{+1.3}  \pm 1.0$   &  $  0.00\pm 0.11 \pm 0.02$  \\
\smallskip
 $\pip \Kz   $    & $172 \pm 17 \pm 9   $     &   $17.5
_{-1.7}^{+1.8}  \pm 1.3$   &  $  -0.17 \pm 0.10  \pm 0.02$  \\
\smallskip
 $\Kp \Kzb  $    & $<10$ ($-5.6_{-5.5}^{+2.8} \pm 2.5$) &
 $<1.3$ ($-0.6_{-0.7}^{+0.6} \pm 0.3$) & $-$  \\  \hline\hline
\end{tabular}
\end{center}
\label{tab:hpi0brresult}
\end{table}

Systematic uncertainties on the branching fractions arise primarily
from uncertainty on the final  selection efficiency   and uncertainty
on $n_i$  due to imperfect knowledge of the PDF shapes.  The latter is
estimated either  by varying the PDF parameters within $1\sigma$ of
their measured uncertainties  or by substituting alternative PDFs from
independent control samples.  In the $h\piz$ analysis the most
relevant systematic uncertainties on the signal yields are due to the
background \mes parametrization and Fisher background shape (about $10\%$
each), while for the $h \KS$ analysis the $\Delta E$ offset and
resolution and Fisher signal shape  contribute the largest
errors (about $4\%$ each). We estimate the systematic uncertainty on the signal yields due to the residual presence of $B$ decay backgrounds with Monte Carlo techniques and we find that it is  negligible compared with the other effects. 
 
In the case of the $\pip \piz$ final state, we evaluate how the
imperfect knowledge of the PDF shapes can affect the significance of
the signal. We recalculate the square root of the change in
$-2\ln{\cal L}$ with $n_{\pip \piz}$ fixed to zero for the worst case
PDF variations and we find a $4.0 \sigma$  statistical
significance for the signal.

Systematic uncertainties on the charge asymmetries are evaluated from PDF variations added in quadrature with the limit on intrinsic charge bias in the detector (0.01).   The small yield of  $\pip \piz$ channel is the origin of the systematic error on the charge asymmetry (0.10),  dominated by the PDF variations.

In conclusion, we find evidence for  the decay $\Bp \to \pip \piz$ and
measure a branching fraction of $\BR(\Bp \to
\pip \piz ) = (4.1^{+1.1}_{-1.0}\pm 0.8)\times 10^{-6}$.  We also
measure $\BR( \Bp \to \Kp \piz ) = (11.1^{+1.3}_{-1.2} \pm 1.0)\times
10^{-6}$ and $\BR( \Bp \to \pip \Kz ) = (17.5^{+1.8}_{-1.7} \pm
1.3)\times 10^{-6}$, with significant improvements on the errors with
respect to our previously published results.  We
do not observe any evidence of direct \CP\ asymmetry in these channels, 
measuring $\acp_{\pip \piz} = 
-0.02 ^{+0.27} _{-0.26} \pm 0.10$, $\acp_{\Kp \piz} =
0.00\pm 0.11 \pm 0.02$, and $\acp_{\pip \Kz} = -0.17 \pm
0.10  \pm 0.02$.  No evidence of a signal is found for the $\Kp \Kzb$
final state for which we set a $90\%$ CL upper limit on the branching
ratio of $1.3 \times 10^{-6}$.

\section{Acknowledgements}
\input{pubboard/acknowledgements}

\end{document}

%% file: symbols.tex
\def\Dp      {\ensuremath{D^+}}
\def\Dm      {\ensuremath{D^-}}
\def\Dstar   {\ensuremath{D^{*}}}
\def\Dstarp  {\ensuremath{D^{*+}}}
\def\Dstarm  {\ensuremath{D^{*-}}}

\newcommand{\etal}{{\it et al.}}

\def\fish    {\ensuremath{\cal F}}

\def\acp {\ensuremath{{\cal A}}}

\def\thsph    {\ensuremath{\theta_{\scriptscriptstyle S}}}

%% file: pubboard/authors_conf02011.tex
\begin{center}
\small

The \babar\ Collaboration,
\bigskip

B.~Aubert,
D.~Boutigny,
J.-M.~Gaillard,
A.~Hicheur,
Y.~Karyotakis,
J.~P.~Lees,
P.~Robbe,
V.~Tisserand,
A.~Zghiche
\inst{Laboratoire de Physique des Particules, F-74941 Annecy-le-Vieux, France }
A.~Palano,
A.~Pompili
\inst{Universit\`a di Bari, Dipartimento di Fisica and INFN, I-70126 Bari, Italy }
% ...

\end{center}\newpage

%% file: pubboard/acknowledgements.tex
We are grateful for the 
extraordinary contributions of our \pep2\ colleagues in
achieving the excellent luminosity and machine conditions
that have made this work possible.
The success of this project also relies critically on the 
expertise and dedication of the computing organizations that 
support \babar.
The collaborating institutions wish to thank 
SLAC for its support and the kind hospitality extended to them. 
This work is supported by the
US Department of Energy
and National Science Foundation, the
Natural Sciences and Engineering Research Council (Canada),
Institute of High Energy Physics (China), the
Commissariat \`a l'Energie Atomique and
Institut National de Physique Nucl\'eaire et de Physique des Particules
(France), the
Bundesministerium f\"ur Bildung und Forschung
(Germany), the
Istituto Nazionale di Fisica Nucleare (Italy),
the Research Council of Norway, the
Ministry of Science and Technology of the Russian Federation, and the
Particle Physics and Astronomy Research Council (United Kingdom). 
Individuals have received support from 
the A. P. Sloan Foundation, 
the Research Corporation,
and the Alexander von Humboldt Foundation.